\documentstyle[aps,prl,multicol,epsfig]{revtex} 

\newcommand{\br}{{\boldmath{r}}}
\newcommand{\ellx}{\ell_\times}
\newcommand{\mW}{{\sf W}}
\newcommand{\mK}{{\sf K}}
\newcommand{\mF}{{\sf F}}
\newcommand{\mone}{{\sf 1}}
\newcommand{\mQ}{{\sf Q}}
\newcommand{\tw}{\widetilde{w}}
\newcommand{\tk}{\widetilde{\kappa}}
\newcommand{\vpsi}{\vec{\psi}}
\newcommand{\vchi}{\vec{\chi}}
\newcommand{\vPhi}{\vec{\varphi}}
\newcommand{\vxi}{\vec{\zeta}}
\newcommand{\talpha}{\widetilde{\alpha}}
\newcommand{\tbeta}{\widetilde{\beta}}

\title{Hydrodynamics of the Kuramoto-Sivashinsky Equation in Two Dimensions}

\author{Bruce Boghosian$^{1}$, Carson C.~Chow$^{2,*}$, 
and Terence Hwa$^{3}$}
\address{$^{1}$ Center for Computational Science, Boston University,
Boston MA 02215.\\
$^{2}$Department of Mathematics, University of Pittsburgh, Pittsburgh
PA 15260.\\
$^{3}$Department of Physics, University of California at San Diego,
La Jolla, CA  92093-0319}

\begin{document}

\maketitle

\begin{abstract}
The large scale properties of spatiotemporal chaos in the 2d
Kuramoto-Sivashinsky
equation are studied using an explicit coarse graining scheme. A set of 
intermediate equations are obtained.  They describe interactions 
between the small scale (e.g., cellular) structures and
the hydrodynamic degrees of freedom. Possible forms of the effective
large scale hydrodynamics are constructed and examined. Although  a number
of different universality classes are allowed by symmetry, numerical
results support the simplest scenario, that being the KPZ universality class.  

\vspace{10pt}
\end{abstract}

\begin{multicols}{2}
\narrowtext

A major goal in the study of spatiotemporal chaos (STC)~\cite{cross93}
is to obtain quantitative connections
between the chaotic dynamics of a system at small scales and 
the apparent stochastic behavior at large scales.
The Kuramoto-Sivashinsky (KS) equation \cite{ksrefs}
\begin{equation}
\partial_t h = -\nabla^2 h - \nabla^4h +(\nabla h)^2  
\label{KS}
\end{equation}
has been used as a paradigm in efforts to elucidate
the micro-macro connections~\cite{yakhot81,zaleski}.  

The qualitative behavior of the KS equation
is quite simple.
Cellular structures are generated at scales of the order $\ell_0 =
2\sqrt{2}\pi$ 
due to the linear instability.  These cells
then interact chaotically with each other via the nonlinear spatial coupling 
to form the STC steady state at scales much larger than $\ell_0$.
The characterization of the STC state has been studied
extensively in one spatial dimension~\cite{yakhot81,zaleski,sneppen92,ch95}.
It was conjectured early by Yakhot~\cite{yakhot81},  based partially
on symmetry grounds, that the large scale
behavior of the one dimensional KS (1d-KS) equation is equivalent to
that of the 1d noisy 
Burgers equation, also known as  the KPZ equation~\cite{kpz}.
This conjecture has since been validated by detailed numerical 
studies~\cite{zaleski,sneppen92}. More recently, 
an explicit coarse-graining procedure was used
by Chow and Hwa~\cite{ch95} to derive a set of
coupled effective equations 
describing the interaction between the chaotic cellular dynamics
and the long wavelength fluctuations of the $h$-field. 
From this description,  the large scale (KPZ-like) behavior of the 1d-KS
system can be predicted quantitatively from the knowledge of various
response functions  
at the `mesoscopic scale' of several $\ell_0$'s.

The behavior of the 2d-KS equation is not as well understood. The simplest
scenario is the generalization of Yakhot's conjecture to 2d,
with the large scale behavior described by the 2d-KPZ
equation,
\begin{equation}
\partial_t h = \nu \nabla^2 h + \frac{\lambda}{2} (\nabla h)^2 +
\eta(\br,t), \label{kpz}
\end{equation}
where $\nu > 0 $ can be interpreted as a stabilizing `surface tension'
for the height profile $h$,
and $\eta$ a {\em stochastic} noise with $\langle
\eta(\br,t)\eta(\br',t')\rangle =2D\delta^2(\br-\br')\delta(t-t')$.  For $\nu
>0$, the asymptotic scaling properties of Eq.~(\ref{kpz}) are
described by `strong-coupling' behavior with algebraic (rather than
logarithmic) scaling in the roughness of $h$
and super-diffusive dynamics~\cite{kpzrev}.
The length scale at which the asymptotic regime is reached is given by
$\ellx \sim e^{8\pi/g}$, where $g\equiv \lambda^2 D/\nu^3$.
At scales below $\ellx$, the effect of
the nonlinear term in (\ref{kpz}) can be accounted for adequately via
perturbation theory. The system behaves in this `weak-coupling' regime
as a linear stochastic
diffusion equation with additive logarithmic corrections~\cite{tang}.

Previous studies of the 2d-KS equation~\cite{proc92a,proc92b,jaya93} 
found behavior consistent with linear
diffusion with logarithmic corrections but had different interpretations.
Jayaprakash {\it et al.}~\cite{jaya93} performed a numerical analysis akin to 
Zaleski's on the 1d-KS~\cite{zaleski}, and concluded that 
their results were consistent with the weak-coupling regime of the 
2d-KPZ equation, with (in principle) a crossover to strong-coupling beyond a
length of $\ellx \approx 10^{26}\ell_0$, for $g=0.4$.
Procaccia {\it et al.}~\cite{proc92a,proc92b} used
a comparative Dyson-Wyld
diagrammatic analysis of the two equations to argue
that 2d-KS and 2d-KPZ {\em cannot} belong to 
the same universality class~\cite{note2}.  
They maintained instead that the asymptotic behavior
of the 2d-KS equation is described by a `nonlocal' solution,
consisting of diffusion with {\em multiplicative} logarithmic 
corrections~\cite{note1}. We feel that the
ensuing debate~\cite{comments} failed to rule out
either interpretation.

It is very difficult to distinguish between the above two scenarios
numerically, as one must
resolve different forms of logarithmic corrections to
the (already logarithmic) correlation function of the linear diffusion
equation. Theoretically, there is no {\it a priori} reason why
simple symmetry considerations such as
Yakhot's should be valid in two and higher dimensions.
Unlike in 1d where there are only scalar density
fluctuations, the 2d case is complicated because three or more large-$k$ modes
can couple and  contribute to low-$k$ fluctuations.
Such nonlocal interactions in $k$ may not be adequately
accounted for in the type of analysis performed in Refs.~\cite{zaleski,jaya93},
which  numerically impose KPZ dynamics  and then test for
self-consistency.

In this study, we perform a systematic symmetry analysis, taking
into account the possibility of large-k  coupling.
Specifically,  we extend the coarse graining procedure
of Ref.~\cite{ch95} to two dimensions to derive a set of coupled
equations 
describing the local arrangement of cells, and study their effect on the
macroscopic dynamics of the $h$-field. The resulting behavior depends crucially
on the small scale arrangement of the cells. In the simplest
case, the strong-coupling 2d-KPZ behavior is recovered. Nevertheless,
more complicated behaviors are allowed if the microscopic
cellular arrangement exhibits {\em spontaneous rotational symmetry breaking}. 
A number of possible scenarios are listed for this case.  To
determine which of the allowed scenarios is selected by the 2d-KS
equation, we performed numerical measurements of the
cellular dynamics at the mesoscopic scale of 4 to 16 $\ell_0$'s.  
Our results disfavor the occurrence of the more exotic
scenarios, leaving the strong-coupling 2d-KPZ behavior as the most
likely possibility.

As in 1d, we coarse grain over a region of size
$L\times L$, where $L$ is several times the typical cellular size
$\ell_0$.  $h(\br,t)$ is separated into fast cellular modes $h_>$
and slow long wavelength modes $h_<$. Inserting
$h(\br,t)=h_<(\br,t)+h_>(\br,t)$ into Eq.~(\ref{KS}), we obtain the following
equations for the fast and slow modes:
\begin{eqnarray}
&\partial_t h_> &= -\nabla^2 h_> - \nabla^4 h_> +(\nabla
h_>)^2_> + 2 (\nabla h_> \cdot \nabla h_<)
\label{fast}\\
&\partial_t h_< &=-\nabla^2h_< + (\nabla h_<)^2 + w(\br,t)
+ O(\nabla^4h_<).
\label{slowdyn}
\end{eqnarray}
where $w(\br,t) \equiv (\nabla h_>)^2_<$ is the only contribution of the 
fast modes on the dynamics of $h_<$. It can be interpreted as  the `drift
rate' of $h_<$ over a regime of $L\times L$ centered at $\br$.  
To specify the dynamics of $h_<$, it is necessary to 
obtain the dynamics of $w$ from the fast mode equation (\ref{fast}).
Due to the structure of the nonlinear term, we must consider
the tensor $\mW$, with elements
$W_{ij}= 2\ \overline{\partial_i h_>\cdot\partial_j h_>}$
where the over-line denotes a spatial average over the coarse-graining
scale $L$.  It is convenient to introduce the curvature tensor
 $\mK$, with elements $K_{ij}= 2 \partial_i\partial_j h_<$.
In this notation, $w= \frac12{\rm Tr}\, \mW$ and
$\kappa \equiv \nabla^2 h_< = \frac12 {\rm Tr}\, \mK$.  
Taking the time derivative of $W_{ij}$ and using
(\ref{fast}), we obtain
\begin{equation}
\partial_t \mW = \mF\left[\mW\right] + \mW\cdot \mK + \mK \cdot \mW
\label{Wdyn}
\end{equation}
where $\mF\left[\mW\right]$ contains purely fast mode dynamics and
will be described shortly. 
The forms of the last two terms in Eq.~(\ref{Wdyn})
are fixed by the Galilean invariance of the KS equation and are exact.

Equation (\ref{Wdyn}) can be made more transparent by
rewriting the two tensors as $\mW = w \cdot \mone + \tw \cdot \mQ(\phi)$ 
and $\mK = \kappa \cdot \mone + \tk \cdot \mQ(\theta)$, where  $\mone$
is the identity matrix and $\mQ(\alpha)$ 
is a unit {\em traceless} matrix, represented by an
angle $\alpha$,
e.g., $Q_{12}(\alpha)=Q_{21}(\alpha)=\sin(2\alpha)$ and 
$Q_{11}(\alpha)=-Q_{22}(\alpha)=\cos(2\alpha)$.
Adopting vector notation $\vpsi= (\tw \cos 2\phi, \tw \sin 2\phi)$
and $\vchi=(\tk\cos 2\theta,\tk\sin 2\theta)$, Eq.~(\ref{Wdyn})
 can be rewritten as
\begin{eqnarray}
\partial_t w&=& f[w] + 2\kappa  w + 2\vchi
\cdot \vpsi \label{wdyn2}\\ \partial_t \vpsi &=&
\vPhi\left[\vpsi\right] + \kappa \vpsi + w \vchi,
\label{psidyn}
\end{eqnarray}
to leading order, with $f$ and $\vPhi$ obtained from the
appropriate decomposition of $\mF$.

Eqs.~(\ref{wdyn2}) and (\ref{psidyn}), together with the slow mode
equation (\ref{slowdyn}), 
form a closed set of coarse grained equations which specifies the
dynamics of $h_<$ once the effective forms of the small scale
dynamics, i.e., $f$ and $\vPhi$ are given. These equations are
constructed from symmetry considerations, and can be regarded as
the more complete generalization of Yakhot's conjecture for two
dimensions.  We first discuss the physical meaning of the coarse grained
variables appearing in $\mW$ and $\mK$.

\begin{figure}
\centerline{
\epsfig{figure=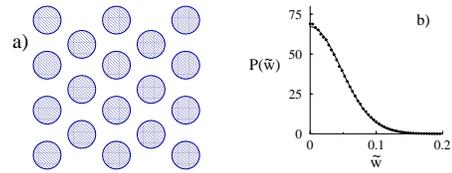,height=2.5in,bbllx=150pt,bblly=100pt,bburx=500pt,bbury=700pt,clip=,angle=270}
}
\caption{a) Example of a cellular arrangement where
$\vpsi$ is nonzero with orientation $2\phi = 90^\circ$, reflecting a
larger `compression' in the $y$ direction. b)
Probability density of 
$\tw=|\vpsi|$. Solid line is a Gaussian fit.}
\label{fig1}
\end{figure}

The tensor $\mK$ describes the local curvature of the slow modes
$h_<$. With $\tk=0$, we have a symmetric paraboloid --- a `valley' if
$\kappa>0$ or a `hill' if $\kappa<0$. With $\kappa=0$, we have a
`saddle', with $\tk$ and $\theta$ specifying the strength and the
orientation. The tensor $\mW$ characterizes the local
{\em packing} of the cells. As in the 1d
case~\cite{ch95}, $w$
gives the local cell density.  The traceless
component of $\mW$ describes the local {\em anisotropy} in
cell packing. 
Fig.~\ref{fig1}(a), shows an example of an arrangement of cells
where $\vpsi$ is pointed in the $y$-direction.  
Fluctuations in the anisotropic part of the curvature $\mK$ 
will affect the local cell packing. For
example, the cell density at the bottom of a valley will be
higher and a saddle configuration in $h_<$ will induce
anisotropy. Eqs.~(\ref{wdyn2}) and (\ref{psidyn}) describe 
these effects of curvature quantitatively, much
like the relation between stress and strain in elastic systems.  Cell
packing in turn influences the slow mode dynamics via the $w$ term in
Eq.~(\ref{slowdyn}). The anisotropic parts of $\mK$ and
$\mW$ are invariant upon a rotation by $180^\circ$; see Fig.~\ref{fig1}(a). 
Thus, we can view
the vector field $\vpsi(\br,t)$ as a `nematic' order parameter
describing the local cellular orientation, and $\vchi$ as an applied
field biasing $\vpsi$ towards a specific orientation.

If we turn off the applied field $\kappa$ and
$\vchi$ in Eqs.~(\ref{wdyn2}) and (\ref{psidyn}), we have $\partial_t
w = f[w]$ and $\partial_t \vpsi = \vPhi\left[\vpsi\right]$ for each
coarse-grained region; thus $f$ and $\vPhi$ describe the small scale dynamics.
Even for a  coarse-grained region of a few $\ell_0$'s, 
the small scale dynamics of $h$ are already
{\em chaotic}. The fields $w(t)$ and $\vpsi(t)$ are `projections'
 of this small scale chaotic dynamics. They can be quantitatively
characterized numerically as we will present shortly. Before doing so,
we first construct some possible scenarios.

We expect that the $h$-field has on average a finite
drift rate, i.e.,  a finite time-averaged value of $w$. 
The simplest dynamics of $w$ is then
$\partial_t w(\br,t)  = f[w] =  -\alpha(w-w_0)  + \xi(\br,t)$,
where $\xi$ is a stochastic forcing mimicking the chaotic small scale
dynamics, and $w_0$ is a constant. This yields $w(\br,t\to\infty) \to w_0$.
The behavior of $\vpsi$ is less straightforward.  
In the simplest scenario,
we have $\partial_t \vpsi = -\talpha \vpsi + \vxi(\br,t)$ to leading order,
with $\vxi$ being a vectorial stochastic forcing.
Eq.~(\ref{psidyn}) then yields (in the hydrodynamic limit)
$\vpsi\simeq (2w_0/\talpha) \vchi$,
where we took the asymptotic result $w=w_0$ 
and assumed that the typical curvature $\kappa$
is small.
Note that in this scenario, the cellular orientation
{\em passively} follows the curvature. In particular, there is no
orientational anisotropy on average if there is no external forcing.
Inserting this result and $f[w]$ into (\ref{wdyn2}), we find in
the hydrodynamic limit
\begin{equation}
w \simeq w_0 +\frac{2w_0}{\alpha}\nabla^2 h_< + \frac{\xi}{\alpha}
+ O((\partial_i\partial_j h_<)^2).
\label{w}
\end{equation}
Substituting (\ref{w}) into (\ref{slowdyn}) yields an equation for $h_<$
of the KPZ form (\ref{kpz}) to leading order, with $\nu = (2w_0/\alpha)-1$ and 
$\eta(\br,t)= \xi(\br,t)/\alpha$. 
Dynamics of the KPZ universality class will be obtained if $\nu > 0$ and the 
noise $\xi$ is uncorrelated between different coarse-graining regions.

Unlike the constant $\alpha$ however, there is no a priori reason
why the constant $\talpha$ cannot be negative. This would be the case
if the microscopic chaotic dynamics has a preference for the
{\em spontaneous} breaking of local isotropy. If $\talpha \le 0$,
then the dynamics of $\vpsi$ would be more complicated.
Higher order terms, e.g., $|\vpsi|^2\vpsi$, will be needed
for stability.  The minimal equation for (\ref{psidyn}) becomes
\begin{equation}
\partial_t \vpsi = -\talpha\vpsi -\tbeta \left|\vpsi\right|^2 \vpsi +
\gamma\nabla^2\vpsi +  w_0 \vchi + \vxi(\br,t)
\label{psifast}
\end{equation}
where $\tbeta$ is a positive constant, and 
the $\gamma$ term describes the coupling of neighboring coarse grained
regions.
Eq.~(\ref{psifast}) describes the relaxational dynamics of a nematic
liquid crystal under an applied `field' $\vchi$.
Its behavior depends crucially on the dynamics of the phase
field, $\phi$, which is the Goldstone mode associated with
symmetry breaking.   
The latter in turn depends on the parameters of Eq.~(\ref{psifast}),
particularly the coupling constant $\gamma$ and the amplitude of the
noise $\vxi$.  The possibilities along with the effects on $h_<$ are:

Case i) If the noise $\vxi$ dominates over the spatial coupling 
$\gamma$, then the local anisotropy will be destroyed at large scales
due to the proliferation of topological defects (disclinations) in
$\phi$. Isotropy is restored and the KPZ universality class is
recovered.

Case ii) If the spatial coupling is large, then the direction of $\vpsi$
 may `phase-lock' with the direction of $\vchi$, as 
manifested by $\langle(\theta-\phi)^2\rangle \ll 1$.  
Solving for the steady state of $w$ in
this case gives
$w\simeq w_0 + \frac{w_0}{\alpha} \kappa + \frac{\tw_0}{\alpha}\tk$,
leading to a slow mode equation which is explicitly {\em not} KPZ-like
since  $\tk = \left[(h_{xx}-h_{yy})^2+4h_{xy}\right]^{1/2}$. (In the
KPZ case, $\tk$ comes in at second order and is presumed irrelevant; see
Eq.~(\ref{w})).

Case iii) For intermediate parameters, there may exist a `spin wave'
phase characterized by
$\langle\phi(\br)\phi(0)\rangle = a \log|\br|$.  Here,
`spin wave' fluctuations would add a long range component to the
effective KPZ noise, since $\langle\cos(\phi(r)-\phi(0))\rangle\sim
r^{-a}$. For sufficiently small $a$, it would yield dynamics that are not in
 the KPZ universality class.

To distinguish between these scenarios, 
we numerical measured the `fast mode' dynamics with simulations on
systems of size $L\times L$ with 
$L$ ranging from 32 to 128 with periodic boundary conditions.  (A single
cell had a length $\ell_0\approx 8.9$.)  We used a
simple spatial 
discretization scheme with an Euler time step of .02.    
We first checked for spontaneous symmetry 
breaking by examining the distribution of $\vpsi$.
We measured $\vpsi$ 
spatially averaged over a system of size $L=32$, sampling at
time intervals of $t=4$ over a total period of $t=4.8\times 10^5$.
 The distribution of $|\vpsi|=\tw$ (normalized by the
cylindrical area) is shown in Fig.~\ref{fig1}(b). The 
Gaussian shape (within $\sim 3\%$), indicates a lack of 
spontaneous symmetry breaking strongly supporting the KPZ
scenario. 

We next  measured the response of $\mW$ to an imposed 
curvature tensor $\mK$ using methods
similar to those described in Ref.~\cite{ch95}. In an STC steady state,
we abruptly turned on a forcing term  in Eq.~(\ref{fast}) of the form
$h_<(x,y)=c k^{-2}\sin k_x x \sin k_y y$, with
$k_x=k_y=\pi/2 L$, for a range of amplitudes $c$.  This
configuration yielded eight separate overlapping regions of
size $L/2\times L/2$: four hill and valley regions, and four
saddle regions with orientations $\theta= \pi/4$ and $\theta =
3\pi/4$. 
To account for the spatial
dependence of the curvature $\mK$ within each forcing region, we
simply took the imposed curvature to be the 
average curvature. This gave
$\kappa=\overline{\nabla^2 h_<} =(4/\pi^2)c$.  We
zeroed the first three Fourier modes of $h_>$ after each time step
to remove the nonlinear slow mode response to
the forcing (see Ref.~\cite{ch95}).

\begin{figure}
\begin{center}
\epsfig{figure=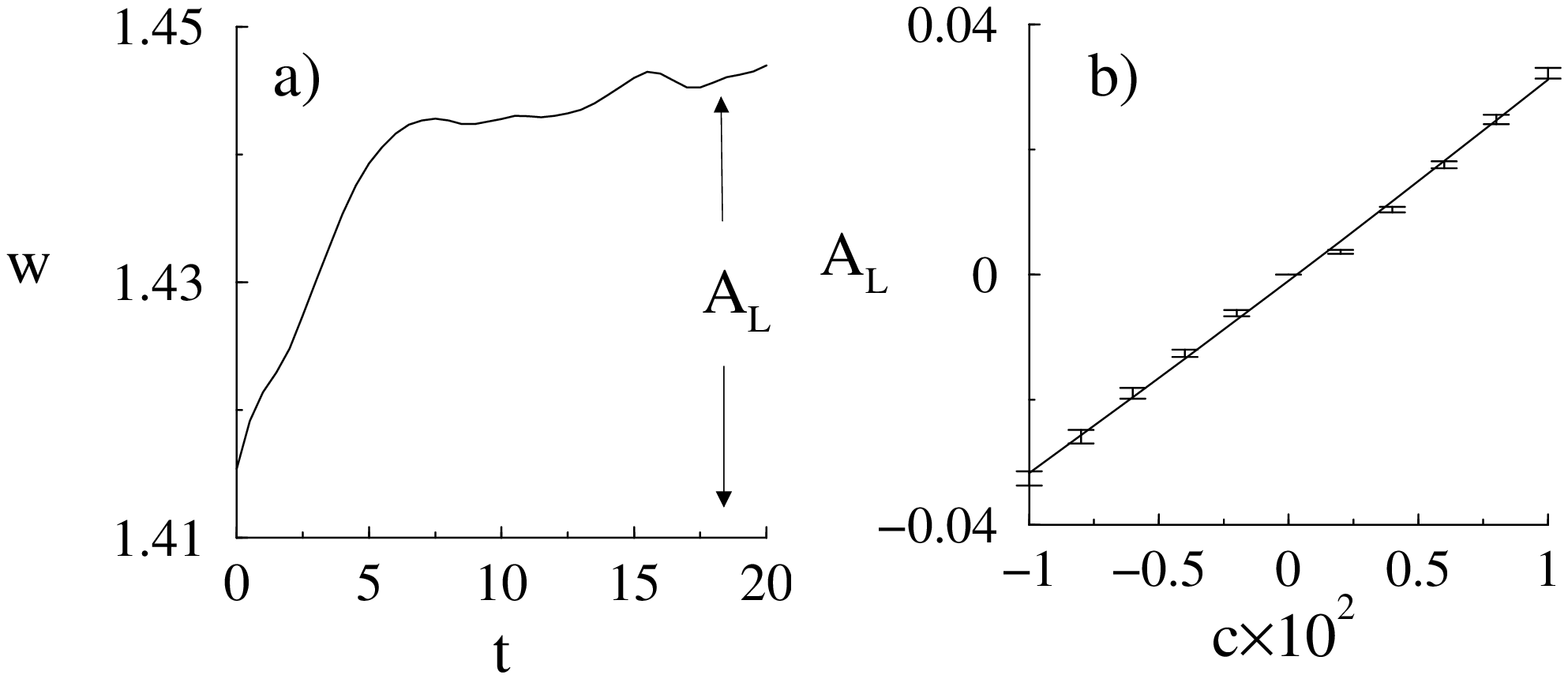,height=2.5in,bbllx=200pt,bblly=100pt,bburx=500pt,bbury=700pt,clip=,angle=270}
\end{center}
\caption{(a) Response of the drift velocity $w$ to forcing with
$c=0.1$ applied at $t=0$ for  $L=128$. 
(b) Saturated amplitude $A_L(c)$;  the slope is $3.15\pm0.06$.
}
\label{fig:scalar}
\end{figure}

We found that $w$ only responded to $\kappa$ and $\vpsi$ only responded
to $\vchi$, further validating the KPZ scenario.
Fig.~\ref{fig:scalar}(a) shows an example of the time dependent response of the
drift rate $w$ to the forcing.  The saturated amplitude $A_L(c)$ was a
linear function 
of $c$ as seen in Fig~\ref{fig:scalar}(b).  In accordance with Eq.~(\ref{w}),
the ratio $2w_0/\alpha$ was computed from the slope, from which we obtained
$\nu=14.9\pm 0.5$ for $L=64$ and
$\nu = 14.5 \pm 0.5$ for $L=128$.    
The averaged response times were 
$\alpha^{-1}= 3.3\pm  0.2$ for $L=64$, 
and  $\alpha^{-1}= 4\pm 0.2$ for $L=128$.
Similar behaviors were obtained in
the hyperbolic forcing region. We verified that
$\vpsi\propto\vchi$, and found that
$\talpha^{-1}= 2.0\pm 0.1$ for $L=64$, and 
$\talpha^{-1}= 3.0\pm 0.3$ for $L=128$.

To characterize the effective stochasticity, we measured the
two-point correlation function  $C_L(\tau)=\langle
(w(t+\tau)- w(t))^2\rangle$ of the drift rate $w$
for systems of sizes ranging from
$L=32$ to $L=128$.  The relevant  quantity is the correlator
$D_L(\tau)=[C_L(\infty)-C_L(\tau)]/2$ (see Fig.~\ref{fig:noise}(a)).  For
short-range correlated noise, the effective noise amplitude 
$D=L^2\int_0^\infty D_L(\tau) d\tau$ is expected
to be independent of $L$.  Fig.~\ref{fig:noise}(b) shows measurements
of $D$, with the average $D=89\pm 5$. 
The numerical values of the effective parameters $D$ and $\nu$ extracted
using our coarse-graining scheme are in reasonable agreement with that
of Ref.~\cite{jaya93}. 
\protect{\begin{figure}
\begin{center}
\epsfig{figure=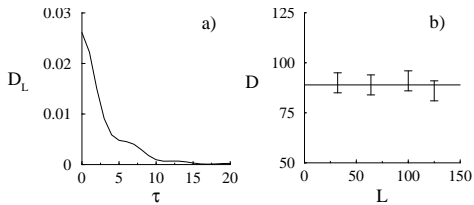,height=2.5in,bbllx=200pt,bblly=100pt,bburx=500pt,bbury=700pt,clip=,angle=270}
\end{center}
\caption{(a) Correlator for $L=32$;
(b) Noise amplitude for different $L$'s.}
\label{fig:noise}
\end{figure}
}

In summary, we have performed an analysis of the 2d-KS equation
at the mesoscopic scale of several cell sizes. By using an explicit 
coarse graining scheme, we constructed various possibilities for the effective
equation of motion for the slow modes $h_<$. We found, as pointed out
before in~\cite{proc92a,proc92b}, that the KPZ universality
class is not the only possibility for the 2d-KS equation. 
Since our analysis is confined 
to the mesoscopic (large-k) limit, the various scenarios obtained are
all nonperturbative and nonlocal in $k$. The more interesting scenarios involve
spontaneous breaking of rotational symmetry. Although we found no
symmetry breaking for the KS equation numerically, our analysis indicates
that such solutions are allowed by symmetry, and may  occur in other 
KS-like system. For the KS equation proper, we conclude that it
belongs to the KPZ universality class. However, for all practical purposes,
the behavior is well described by a stochastic diffusion equation with 
logarithmic corrections.

This research is supported in part by NIH grant K01 MH1058
(CC), ONR grant  N00014-95-1-1002 (TH), and the A.P.~Sloan
Foundation (CC,TH).

\end{multicols}


\begin{thebibliography}{99}
\bibitem[*]{cc} To whom correspondences should be addressed.
\bibitem{note2} Their argument is contingent upon an equality (Eq.~(23) of
Ref.~\cite{proc92a}) relating the difference of two integration constants 
(called $C_1$ and $C_2$) and the bare coefficient of the diffusion term.
The nonlocal solution is tenable if the equality is satisfied.
They claim that the constants, which come
from Wyld diagrammatic calculations, are determined {\em uniquely} 
by the $(\nabla h)^2$nonlinearity and therefore the {\em same}
for both the KS and KPZ equations. However,
the calculation involved integration of
the `full' response and correlation functions
over the {\em entire} range of $k$. We note that
these functions should be
different for large values of $k$'s where the microscopic
dynamics matter. Consequently the constants $C_1$ and $C_2$ need {\em not}
 be the same for the KS and KPZ equations.
\bibitem{note1} The nonlocal solution of Ref.~\cite{proc92a}
has the same form as that of the $g=0$ fixed point of the 2d-KPZ equation.
For the latter, multiplicative logarithmic correction to diffusion
arises from the `asymptotic freedom' of the system as $g \to 0^-$.
The conjectured exponents of the logarithmic correction factor
follow naturally from the nonrenormalization of $\nu$ of the 
KPZ equation to all orders in $\lambda$.
\end{thebibliography}
\end{document}